\documentclass[preprint]{rsl}
\title{Dense Gas History of the Universe: from ASPECS to the ngVLA}
\author{C.L. Carilli, F. Walter, R. Decarli, M. Aravena, Dominik A. Riechers, J. Gonzalez-Lopez, Yali Shao, L. Boogaard, R. Bouwens, M. Neeleman}

\begin{document}

\maketitle

%
% Manuscript contents begins.
%

\begin{abstract}
We review the evolution of the cosmic average molecular gas density to large look-back times, using observations of rotational transitions of CO. Molecular gas is the fuel for star formation in galaxies. Deep searches for CO emission from distant galaxies have delineated the density of molecular gas back to $z \sim 5$, or within 1~Gyr of the Big Bang. The results show a rise and fall in the gas density that parallels, and likely drives, the rise and fall of the cosmic star formation rate density. We present the potential for the next generation Very Large Array to image the distribution and dynamics of the molecular gas in early galaxies, and to make a precise measurement of the dense gas history of the  Universe.

\end{abstract}

\section{Introduction}

Radio astronomy plays key roles in studies of galaxy formation in the distant Universe ($z > 1$, or t$_{univ} < 6$~Gyr), in many ways\footnote{This paper is synthesized from an invited review on radio observations of galaxy formation presented at the URSI General Assembly in Rome, 2021.}. The primary contributions in the last decade include, as a function of observing frequency: 

\begin{itemize} 
\item At frequencies $< 1$ GHz: observations of HI 21 emission and absorption, using primarily the Giant Meterwave Radio Telescope\footnote{
{https://www.gmrt.org/}}, and more recently, ASKAP\footnote{
 {https://www.atnf.csiro.au/projects/askap/index.html}}, and MeerKAT\footnote{ {https://www.sarao.ac.za/gallery/meerkat/}}, and in the future, the Square Kilometer Array (SKA)\footnote{{https://www.skatelescope.org/}}, determine the extent, temperature, and cosmic mass density of the neutral atomic gas across cosmic time \cite{kanekar}. 

\item At 1~GHz to 10 GHz: observations of thermal Free-Free and non-thermal Synchrotron continuum emission with $\mu$Jy sensitivity surveys, using the Jansky Very Large Array (VLA)\footnote{{https://science.nrao.edu/facilities/vla}}, ASKAP, MeerKat, and in the future, the next generation VLA (ngVLA)\footnote{{https://ngvla.nrao.edu/}}, and SKA, determine dust obscuration-free star formation rates, and radio AGN demographics \cite{morrison,murphy,algera}. 

\item At 30~GHz to 300~GHz: observations of cool molecular gas via the rotational transitions of CO, using the VLA, the Northern Extended Millimeter Array (NOEMA)\footnote{{https://www.iram-institute.org/EN/noema-project.php}}, the Australia Telescope Compact Arrray\footnote{{https://www.narrabri.atnf.csiro.au/}}, and the Atacama Large Millimeter Array (ALMA)\footnote{{https://www.almaobservatory.org/en/home/}} \cite{carilliwalter, tacconi}, determine the mass of the dense gas that acts as the immediate fuel for star formation in galaxies. 

\item At $> 100$~GHz: observations of thermal emission from warm dust, with ALMA and NOEMA, and single dish telescopes, reveal the dust-obscured star formation in early galaxies \cite{aravena, casey}.

\item At $> 250$~GHz: observations of [CII] 158~$\mu$m, and other fine structure lines with ALMA, at $z= 3$ to 10, reveal the atomic gas, and galaxy dynamics back to the first galaxies in the Universe \cite{alpine,rebels}. 
\end{itemize}

In this short contribution, we focus on the latest results for the evolution of the molecular gas density, as measured in the 30 GHz to 300 GHz range, particularly by ALMA with the ASPECS survey. We then discuss the potential for the ngVLA to advance these studies, in the age of high precision cosmology. 

\section{ASPECS Survey}

Molecular gas is the immediate fuel for star formation in galaxies, and hence constitutes a key constituent in models of galaxy formation \cite{carilliwalter, tacconi}. Only recently, measurements of the cosmic volume average density of molecular gas have been made, through the advent of  large, wideband interferometers, such as the Jansky Very Large Array, the Atacama Large Millimeter Array and NOEMA. While the molecular gas mass is dominated by $H_2$, for practical reasons, a primary (although not exclusive), method for tracing molecular gas mass entails observation of the rotational transitions of CO, and adoption of calibrated conversion factors \cite{bollato}.

\begin{figure}
\includegraphics[width=\columnwidth]{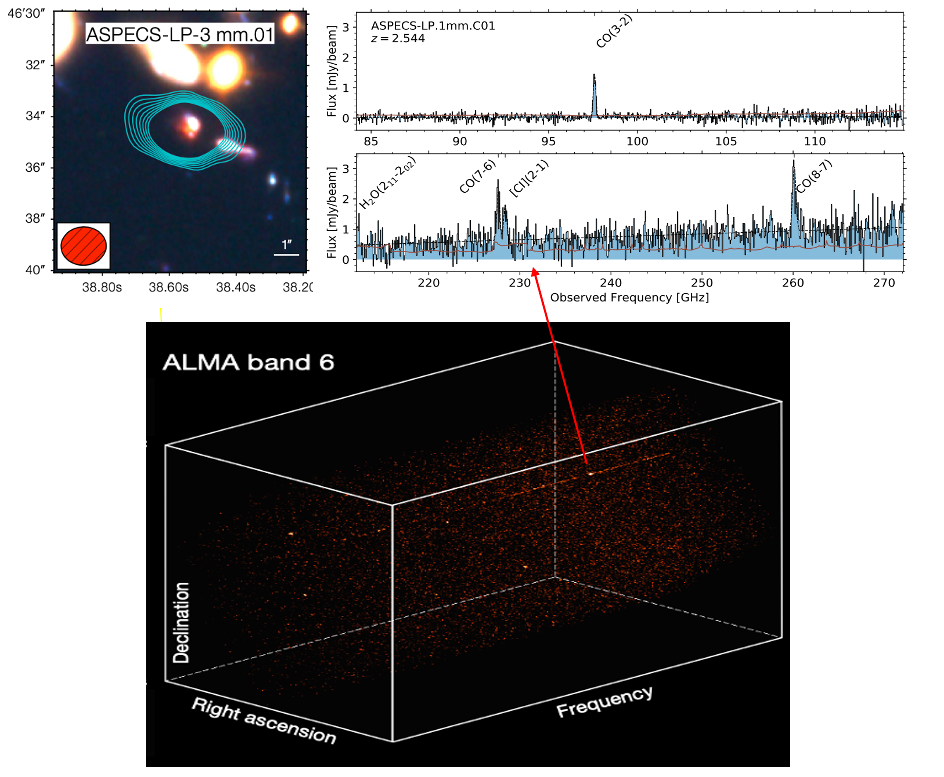}
  \caption{A 3D view of the ASPECS survey (RA, Dec, Frequency = distance), in the ALMA 240~GHZ bands, and an example image and spectra of one galaxy \cite{aravena2,gonzalez19,boogard19}. The bright knots are emission lines, and the broad streak indicates a thermal dust continuum signal.}
  \label{fig:Field}
\end{figure}

The ALMA Spectroscopic Survey (ASPECS), is the definitive deep blind search for molecular gas via CO emission in the distant Universe \cite{walter16}. The survey covered 5 arcmin$^2$ in the Hubble Ultra-Deep field, scanning the full ALMA 90~GHz and 230~GHz bands at $\sim 1"$ resolution, covering multiple CO transitions down to gas mass limits of a few 10$^9$~M$_\odot$, and covering a comoving volume of 42,000 Mpc$^3$. Figure~\ref{fig:Field} shows a three-dimensional view of the ASPECS field (two sky coordinates, and frequency = redshift = distance), with an example of a CO detected galaxy \cite{aravena,walter16,gonzalez19}. A total of 32 CO emission lines were detected from galaxies at $z=0.5$ to 3.6. 

Properties of this unique galaxy sample selected by molecular gas mass include:

\begin{itemize}
    \item 60\% are detected in dust continuum in the associated deep ALMA image, with star formation rates $> 10$ M$_\odot$ year$^{-1}$ \cite{aravena}.
    \item 100\% are detected in the deep optical and near-IR images of the UDF \cite{boogard19,aravena}.
    \item 70\% are `main sequence' disk galaxies, and the rest are compact or irregular \cite{boogard19}.
    \item Metallicities are typically Solar or greater \cite{boogard19}. 
    \item Chandra X-ray observations imply 20\% of the galaxies host AGN \cite{boogard19}. 
    \item Multi-line CO observations imply modest excitation \cite{boogard20}.
\end{itemize}

A key result from ASPECS is the confirmation that the typical gas mass to stellar mass ratio in massive disk galaxies increases from $\sim 0.1$ in the nearby Universe, to $\ge 1$ at $z \sim 3$ \cite{aravena19}. Also, the deep 230 GHz continuum image from the survey implies a clear flattening of the source counts below 0.1 mJy, which has important implications for future submm deep fields \cite{gonzalez20}.

\begin{figure}
  \includegraphics[width=\columnwidth]{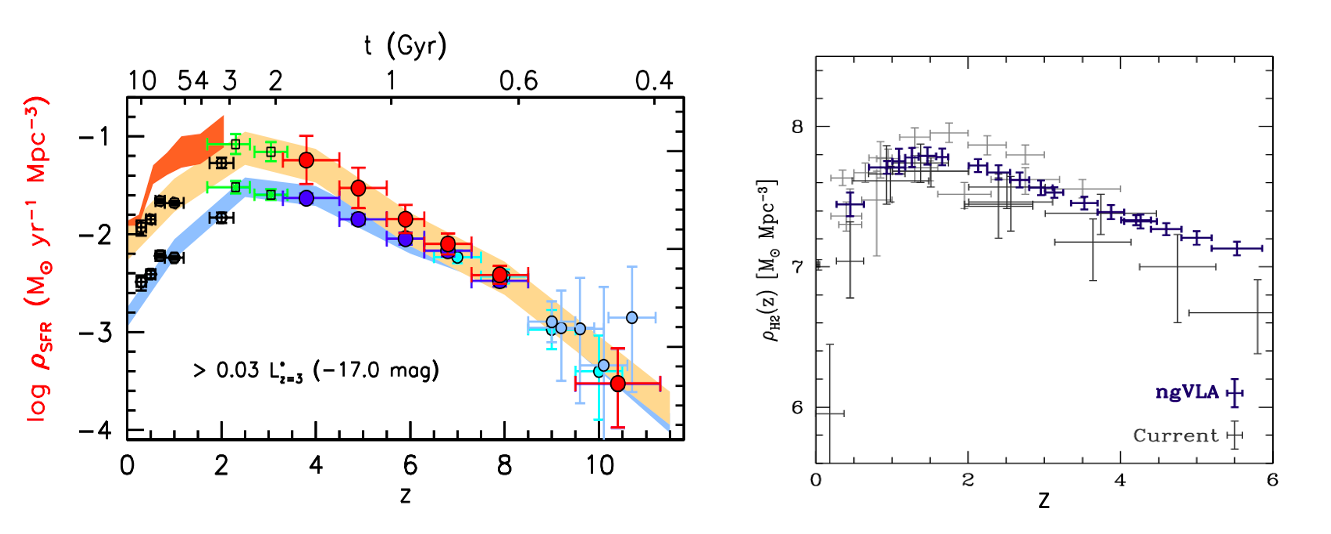}
  \caption{Left: Evolution of the cosmic volume average density of star formation rate \cite{bouwens}. The upper points, highlighted by the orange curve, include a correction for dust obscuration, while the lower blue-highlighted curve does not.
  Right: Evolution of the cosmic volume average molecular gas density \cite{decarli,riechers,walter20,walter20b}. The black and grey points are current measurements, and the blue points and errors are predictions for the ngVLA deep fields. 
  }
  \label{fig:Rho}
\end{figure}

The ASPECS survey, along with other deep blind surveys, such as the VLA COLDZ large program \cite{riechers}, and targeted observations of known galaxies \cite{tacconi}, have determined the evolution of the cosmic volume average density of molecular gas to within $\sim 1$~Gyr of the Big Bang, or a look-back time of $\sim 13$~Gyr. Figure~\ref{fig:Rho} shows a compendium of molecular gas measurements, along with the evolution of the cosmic star formation rate density \cite{bouwens}. Both quantities show a rise and fall with cosmic time, peaking in the range $z \sim 1.5$ to 2.5 (t$_{univ}$ = 3.3~Gyr). This peak in the cosmic star formation rate density, during which about half of the stars in the current Universe form, has long been known \cite{madau,driver}, but not fully explained. The parallel rise and fall in the molecular gas density provides strong circumstantial evidence that the evolution of the cosmic star formation rate density is a consequence of the evolution of the molecular gas content of galaxies. 

The molecular gas measurements were the last piece in the puzzle of the evolution of the baryonic matter associated with galaxies. A consistent picture is emerging in which galaxies accrete ionized material from the intergalactic medium, which passes through a neutral atomic phase, then builds up as the molecular phase, which then collapses to from stars \cite{walter20}. The need for gas replenishment to fuel continued star formation in early galaxies was pointed out in the early paper by \cite{hopkins}. The latest measurements confirm this requirement. 

Unfortunately, the molecular gas measurements remain limited, with large errors at each redshift, and the ability to perform high resolution imaging of the gas in distant galaxies remains prohibitively expensive, even with the biggest existing interferometers. 

\section{Next Generation Very Large Array}

The next generation Very Large Array (ngVLA) constitutes an order of magnitude increase in collecting area over the current VLA and ALMA, covering the frequency range from $\sim 1$~GHz to 116~GHz, with baselines from tens of meters to thousands of kilometers \cite{selina,murphy18}. A precise measurement of the dense gas history, and high resolution imaging of molecular gas in distant galaxies, are primary science drivers for the ngVLA \cite{decarli18,carillishao}. In just a few hours integration, the ngVLA can detect a  gas mass of $\sim 10^9$~M$_\odot$ at $z \sim 2$, at a resolution of $0.15''$. The wide frequency range of the ngVLA also allows for multi-transition studies of molecular gas excitation, including the low order transitions which provide perhaps the best measure of total molecular gas mass \cite{vlaspecs}.

In terms of the cosmic density of molecular gas, Figure~\ref{fig:Rho} shows the capability of the ngVLA vs. the best current measurements. The increased bandwidth and sensitivity of the ngVLA will increase the survey speed by almost two orders of magnitude, revealing thousands of galaxies in blind surveys (as opposed to the current tens). The results will provide a precise measurement of the evolution of the molecular mass density \cite{decarli18}, to complement future measurements of the cosmic star formation rate density by JWST and the ELTs.

\begin{figure}
  \includegraphics[width=\columnwidth]{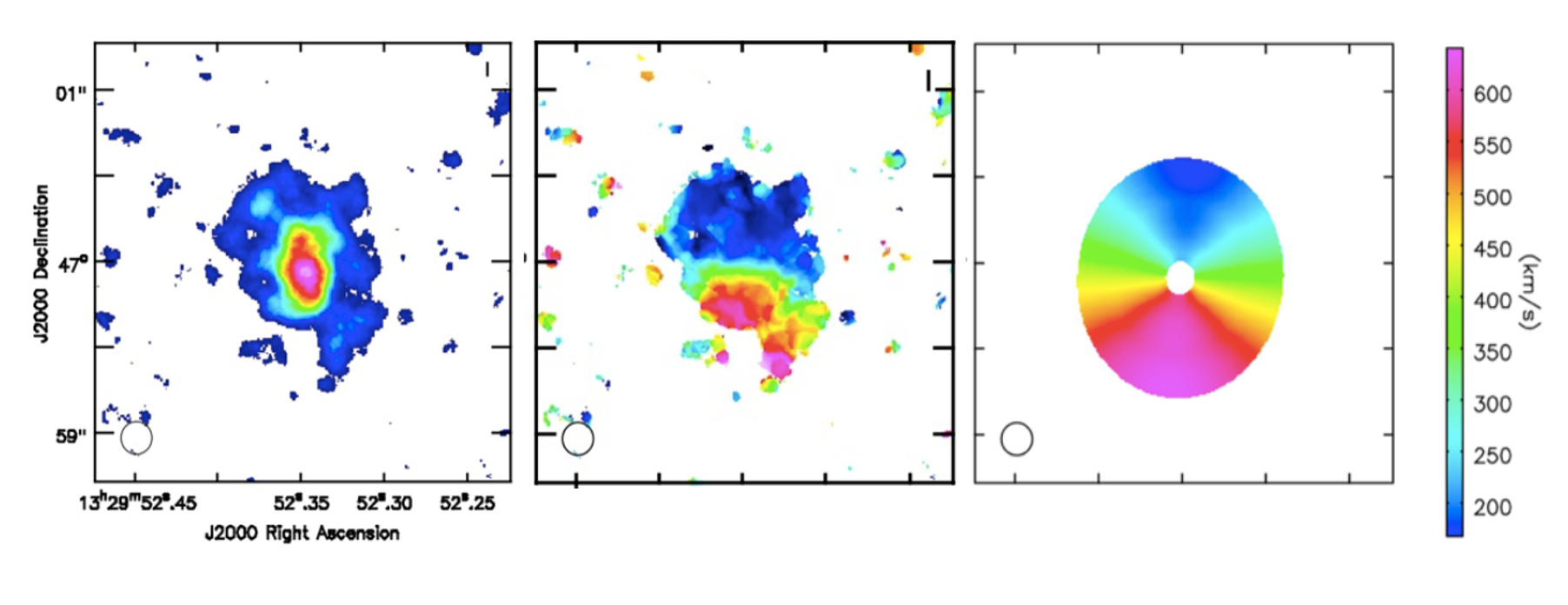}
  \caption{Simulation of an ngVLA observation at 44~GHz of a $z = 4.2$ massive disk galaxy. Left is the total molecular gas column density, center is the intensity weighted mean velocity, and right is a best-fit model of a rotating disk galaxy \cite{carillishao}.}
  \label{fig:M51}
\end{figure}

For CO imaging, Figure~\ref{fig:M51} shows a simulation of ngVLA observations (30 hours) of CO 2-1 emission from a massive disk galaxy at $z = 4.2$ at a resolution of $0.2"$ (total gas mass $\sim$ few$\times10^{10}$~M$_\odot$), including velocity integrated CO emission (column density), and the mean velocity. The gas can be easily traced over the 10~kpc disk, and the velocities are well determined. Also shown is a rotating disk model fit to the data, from which a rotational velocity, and even radial profile, can be derived, potentially determining the dark matter content of the first galaxies. Such  observations are well beyond the capability of existing facilities \cite{carillishao}. 

\begin{figure}
  \includegraphics[width=\columnwidth]{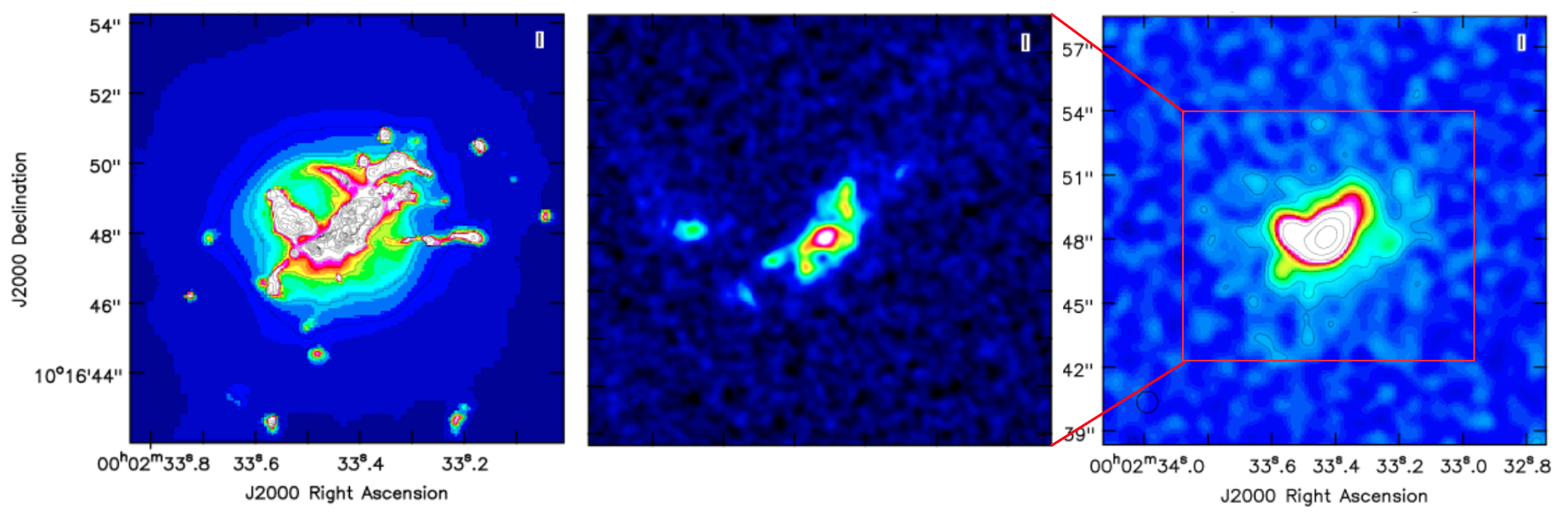}
  \caption{Simulation of an ngVLA observation at 38~GHz of CO 1-0 emission from a $z = 2$ forming giant galaxy and proto-cluster. Left is the model. Center is the total molecular gas column density at $0.13"$ = 1~kpc resolution for the simulated observations. Right is the column density at $1"$ resolution (adapted from \cite{carillierickson}).}
  \label{fig:spider}
\end{figure}

Figure~\ref{fig:spider} shows an ngVLA simulated observation of CO 1-0 emission of a forming massive cluster of galaxies at $z = 2$. A recent exciting discovery has been the detection of CO emission on tens to 100 kpc-scales in forming clusters of galaxies \cite{emonts}. The ngVLA has the ability to resolve the molecular gas distribution down to 1~kpc resolution, and to delineate the extended CO emission out to $\sim 100$ kpc \cite{emonts18}. 

\section{Discussion: Future Context}

We have reviewed the current status of measurements of the evolution of the molecular gas content of galaxies back to within $\sim 1$Gyr of the Big Bang, and it's relationship to the star formation history of the Universe. We then present the capabilities of the next generation VLA to advance these studies into the next decade, to obtain a precise measurement of the dense gas history of the Universe. We briefly discuss the broader radio astronomy landscape in the immediate future, and beyond. 

A number of telescopes arrays are either starting operation, or expected to attain first light in the coming 5 years. These include the current operations of the $\mu$GMRT, MeerKAT, and ASKAP, and the future operation of phase I of the SKA. Regardless of sensitivity, all of these facilities operate at frequencies below 15 GHz. Hence, while they are, or will be, powerful devices to study the HI 21cm emission and radio continuum emission from distant galaxies, they will be unable to study the primary rotational molecular transitions, except for CO 1-0 at $z  > 7$. At such high redshift, this transition is highly suppressed due to depopulation of the lower excitation states by the CMB, and effectively undetectable. The SKA phase II may eventually go up to 30 GHz, still being limited to only CO 1-0 at high redshift ($z > 3$).

Study of the rotational molecular transitions remains the territory of millimeter telescopes, such as the current NOEMA and ALMA, and, toward the end of the decade, the ngVLA. We have shown that, in terms of measuring the total gas masses, and imaging of the molecular gas on kpc-scales, the ngVLA, with its order of magnitude improvement in collecting area, will open unique parameter space critical for advancing our understanding of galaxy formation.

%
% Note that the authors' affiliations and contact information should be
% here, after the references.
%
\noindent\small
Chris Carilli, NRAO, Socorro, NM, USA, 87801, ccarilli@nrao.edu\\
Fabian Walter and Marcel Neeleman, Max-Planck Institute for Astronomy, Heidelberg, DE, walter@mpia.de, neeleman@mpia.de\\
R. Decarli, INAF - Osservatorio di Astrofisica, Bologna, Italy, reberto.decarli@inaf.it\\
M. Aravena and J. Gonzalez-Lopez, Facultad di Ingenieria y Ciencias, Univ. Diego-Portales, Santiago, Chile, manuelaravenaa@mail.udp.cl, jorge.gonzalez@mail.udp.cl \\
D. Riechers, Physics Institute, Univ. of Cologne, Cologne, Germany, riechers@ph1.uni-koeln.de \\
Yali Shao, Max-Planck Institute for Radio Astronomy, Bonn, Germany, yshao@mpifr-bonn.mpg.de \\
R. Bouwens, L. Boogaard (Leiden Observatory, Leiden University, Leiden, The Netherlands, bouwens@strw.leidenuniv.nl, boogaard@strw.leidenuniv.nl) \\

\begin{thebibliography}{9}
  
\bibitem{kanekar} A. Chowdhury, N. Kanekar, J. Chengalur et al., ''H I 21-centimetre emission from an ensemble of galaxies at an average redshift of one,'' \emph{Nature}, 
\textbf{586}, Oct 2020  pp. 369-372

\bibitem{morrison} G. Morrison, F. Owen, M. Dickinson, et al. , ''Very Large Array 1.4 GHz Observations of the GOODS-North Field,'' \emph{Astrophysical Journal Supplement}, \textbf{188}, May 2010, pp. 178-186

\bibitem{murphy} E.J. Murphy, E. Momjian, J.J. Condon  et al., ''The GOODS-N Jansky VLA 10 GHz Pilot Survey: Sizes of Star-forming $\mu$Jy Radio Sources,'' \emph{Astrophysical Journal}, 
\textbf{839}, Oct 2020  pp. 35-52
 
\bibitem{algera}  H. Algera, D. van der Flugt, J. Hodge, et al., ''A Multiwavelength Analysis of the Faint Radio Sky (COSMOS-XS),'' \emph{Astrophysical Journal}, \textbf{903}, Nov 2020  pp. 139-166

\bibitem{carilliwalter} C.L. Carilli, F. Walter, ``Cool Gas in High-Redshift Galaxies,''  \emph{Annual Reviews of Astronomy and Astrophysics}, \textbf{51}, Aug. 2013, pp. 105-161

\bibitem{tacconi} L. Tacconi, R. Genzel, A. Sternberg ``The Evolution of the Star-Forming Interstellar Medium Across Cosmic Time,''  \emph{Annual Reviews of Astronomy and Astrophysics}, \textbf{58}, Aug. 2020, pp. 157-203

\bibitem{aravena} M. Aravena, L. Boogaard, J. Gonzalez-Lopez, et al., ``The ALMA Spectroscopic Survey in the Hubble Ultra Deep Field: The Nature of the Faintest Dusty Star-forming Galaxies,''  \emph{Astrophysical Journal}, \textbf{901}, Sept. 2020, pp. 79-91
 
\bibitem{gonzalez19} J. Gonzalez-Lopez, M. Novak, R. Decarli et al., ''The Atacama Large Millimeter/submillimeter Array Spectroscopic Survey in the Hubble Ultra Deep Field: CO Emission Lines and 3 mm Continuum Sources,'' \emph{Astrophysical Journal}, \textbf{892}, July 2019,  pp. 139-160.
  
\bibitem{casey} C.M. Casey, D. Narayanan, A. Cooray, ``Dusty star-forming galaxies at high redshift,''  \emph{Physics Reports}, \textbf{541}, Aug. 2014, pp. 45-161

\bibitem{rebels} R. Bouwens, R. Smit, S. Schouws et al. ``Reionization Era Bright Emission Line Survey: Selection and Characterization of Luminous,'' arXiv:2106.13719, June 2021.

\bibitem{alpine} O. Le Fevre, M. Bethermin, A. Faisst, et al. ``The ALPINE-ALMA [CII] survey. Survey strategy, observations, and sample properties of 118 star-forming galaxies at $4 < z < 6$,''  \emph{Astronomy \& Astrophysics}, \textbf{643}, Nov. 2020, pp. 1-20

\bibitem{bollato} A.D. Bolatto, M. Wolfire, A.K. Leroy, ``The CO-to-H2 Conversion Factor,' \emph{Annual Reviews of Astronomy and Astrophysics}, \textbf{51}, Aug. 2013 pp. 207-268.

\bibitem{walter16} F. Walter, R. Decarli, M. Aravena,
et al., ``The ALMA Spectroscopic Survey in the HUDF: Survey Description,''  \emph{Astrophysical Journal}, \textbf{833}, Dec. 2016, pp. 67-82

\bibitem{boogards} L.A. Boogaard, R.J. Bouwens, D. Riechers et al., ``Measuring the Average Molecular Gas Content of Star-forming Galaxies at z = 3-4,' \emph{Astrophysical Journal}, \textbf{916}, July 2021,  pp. 12-24.

\bibitem{boogard19} L.A. Boogaard, R. Decarli, J. Gonzalez-Lopez et al., ``The ALMA Spectroscopic Survey in the HUDF: Nature and Physical Properties of Gas-mass Selected Galaxies Using MUSE Spectroscopy,' \emph{Astrophysical Journal}, \textbf{882}, Sept 2019,  pp. 140-164.

\bibitem{aravena19} M. Aravena, R. Decarli, J. Gonzalez-Lopez et al., ``The ALMA Spectroscopic Survey in the Hubble Ultra Deep Field: Evolution of the Molecular Gas in CO-selected Galaxies,'' \emph{Astrophysical Journal}, \textbf{882}, Sept 2019,  pp. 136-153

\bibitem{aravena2} M. Aravena, C. Carill, R. Decarli, et al., ``The ASPECS Survey: An ALMA Large Programme Targeting the Hubble Ultra-Deep Field,'' \emph{The Messenger}, \textbf{179}, March 2020, pp. 17-23

\bibitem{boogard20} L.A. Boogaard, P. van der Werf, A. Weiss et al., ``The ALMA Spectroscopic Survey in the Hubble Ultra Deep Field: CO Excitation and Atomic Carbon in Star-forming Galaxies at z = 1-3,'' \emph{Astrophysical Journal}, \textbf{902}, Oct 2020,  pp. 109-138.

\bibitem{gonzalez20} J. Gonzalez-Lopez, M. Novak, R. Decarli et al., ''The ALMA Spectroscopic Survey in the HUDF: Deep 1.2 mm Continuum Number Counts,'' \emph{Astrophysical Journal}, \textbf{897}, July 2020,  pp. 91-101.
  
\bibitem{riechers} D.A. Riechers, R. Pavesi, C.E. Sharon, et al., ``COLDZ: Shape of the CO Luminosity Function at High Redshift and the Cold Gas History of the Universe,''  \emph{Astrophysical Journal}, \textbf{872}, Feb. 2019, pp. 7-31

\bibitem{bouwens} R. Bouwens, J. Gonzalez-Lopez, M. Aravena, et al. 
F. Walter, R. Decarli, M. Aravena et al., ``The ALMA Spectroscopic Survey Large Program: The Infrared Excess of z = 1.5-10 UV-selected Galaxies and the Implied High-redshift Star Formation History,''  \emph{Astrophysical Journal}, \textbf{902}, Oct. 2020, pp. 112-31

\bibitem{driver} S. Driver, S. Andrews, E. da Cunha, et al., ``GAMA/G10-COSMOS/3D-HST: the 0 < z < 5 cosmic star formation history, stellar-mass, and dust-mass densities,''  \emph{Monthly Notices of Royal Astronomical Society}, \textbf{475}, Apr. 2018, pp. 2891-2935

\bibitem{madau} P. Madau, M. Dickinson,``Cosmic Star Formation History,''  \emph{Annual Reviews of Astronomy and Astrophysics}, \textbf{52}, Aug. 2014, pp. 415-486

\bibitem{decarli} R. Decarli, F. Walter, J. Gonzalez-Lopez, et al., ``The ALMA Spectroscopic Survey in the HUDF: CO Luminosity Functions and the Molecular Gas Content of Galaxies through Cosmic History,''  \emph{Astrophysical Journal}, \textbf{882}, Sept. 2019, pp. 138-155

\bibitem{walter20} F. Walter, C. Carilli, M. Neeleman et al. ``The Evolution of the Baryons Associated with Galaxies Averaged over Cosmic Time and Space,''
\emph{Astrophysical Journal}, \textbf{902},  Oct. 2020, pp. 111-122

\bibitem{hopkins} A. Hopkins, M. McClure-Griffiths, B. Gaensler ``Linked Evolution of Gas and Star Formation in Galaxies Over Cosmic History,'' 
\emph{Astrophysical Journal Letters}, \textbf{682},  July 2008, pp. 13-18

\bibitem{walter20b} F. Walter, C. Carilli, R. Decarli et al. ``The evolution of the cosmic molecular gas density,'' \emph{Bulletin of the American Astronomical Society}, \textbf{51},  May 2019, pp. 442-449

\bibitem{murphy18} E. Murphy, A. Bolatto, S. Chatterjee, et al. ``The ngVLA Science Case,''  Science with a Next Generation Very Large Array, ASP Conference Series, Vol. 517, Dec. 2018, pp. 3-14.

\bibitem{selina} R. Selina, E. Murphy, M. McKinnon et al. ``Imaging Molecular Gas at High Redshift,''  Science with a Next Generation Very Large Array, ASP Conference Series, Vol. 517, Dec. 2018, pp. 15-36.

\bibitem{vlaspecs} D. Riechers, L. Boogaard, R. Decarli et al. ''VLA-ALMA Spectroscopic Survey in the Hubble Ultra Deep Field (VLASPECS),'' \emph{Astrophysical Journal Letters}, \textbf{896}, June 2020, pp. 21-29

\bibitem{decarli18} R. Decarli, C.L. Carilli, C. Casey et al., ``Cold Gas in High-z Galaxies: The Molecular Gas Budget,''  Science with a Next Generation Very Large Array, ASP Conference Series, Vol. 517, Dec. 2018, pp. 565-572.

\bibitem{carillishao} C.L. Carilli \& Yali Shao, ``Imaging Molecular Gas at High Redshift,''  Science with a Next Generation Very Large Array, ASP Conference Series, Vol. 517, Dec. 2018, pp. 535-545.

\bibitem{emonts} B.H. Emonts, M. Lehnert, M. Villars-Martin et al. ``Molecular gas in the halo fuels the growth of a massive cluster galaxy at high redshift,''  \emph{Science}, \textbf{354}, Dec. 2016, pp. 1128-1130

\bibitem{emonts18} B.H. Emonts, C.L. Carilli, D. Narayanan, ``Imaging Molecular Gas at High Redshift,''  Science with a Next Generation Very Large Array, ASP Conference Series, Vol. 517, Dec. 2018, pp. 587-594.

\bibitem{carillierickson} C.L. Carilli \& A. Erickson, ``Initial Imaging Tests of the Spiral Configuration,''  Next Generation Very Large Array Memo Series 41, March, 2018 

\end{thebibliography}
\end{document}